\documentclass[preprint,nofootinbib]{revtex4}

\usepackage{graphicx}
\usepackage{amsmath}
\usepackage{caption}
\usepackage{subcaption}
\usepackage{color}

\begin{document}

\title{Non-singular naked solutions in quantum spacetime}

\author{I. P. R. Baranov$^{1}$, H. A. Borges$^{2}$, F. C. Sobrinho$^{3}$ and S. Carneiro\footnote{Corresponding author}\footnote{saulo.carneiro.ufba@gmail.com}$^{2,4}$}

\affiliation{$^1$Instituto Federal de Educa\c c\~ao, Ci\^encia e Tecnologia da Bahia, 40301-015, Salvador, BA, Brazil\\$^2$Instituto de F\'{\i}sica, Universidade Federal da Bahia, 40210-340, Salvador, BA, Brazil\\$^3$Instituto de F\'isica, Universidade de S\~ao Paulo, 05508-090, S\~ao Paulo, SP, Brazil
\\$^4$Observat\'orio Nacional, 20921-400, Rio de Janeiro, RJ, Brazil}

\date{\today}

\begin{abstract}
Polymer models have been used to describe non-singular quantum black holes, where the classical singularity is replaced by a transition from a black hole to a white hole. In a previous letter, in the context of a uni-parametric model with asymptotic flat exterior metric, we fixed the radius of the transition surface through the identification of its area with the area gap of Loop Quantum Gravity. This revealed a dependence of the polymerisation parameter on the black hole mass, where the former increases as the latter decreases, and it also enabled the extension of the model to Planck-scale black holes. We have identified the existence of limiting states with masses $m \geq \sqrt{2}/4$ and zero surface gravity, showing that Hawking evaporation asymptotically leads to remnant black holes of Planck size. In the present paper we consider solutions with $m < \sqrt{2}/4$, observing again the presence of a minimal radius, but without formation of horizons. Diversely from the previous mass range, only charged solutions are allowed in this case.
\end{abstract}

\maketitle

\section{Introduction}

``Mass without mass" and ``charge without charge" are old dreams according to which fundamental particles are, ultimately, nothing than configurations of spacetime. A physical world in which spacetime is not only a scenario, but conforms the actors at the same time \cite{Wheeler}. The discovery of other fundamental interactions than gravity and electromagnetism has certainly turned such dreams harder to realise. Nevertheless, the concept of ``mass without mass" has found possible realisations in the realm of quantum gravity theories, where the quantum fluctuations of spacetime can source the formation of non-singular black holes.

An interesting and relatively recent example can be found in the context of Loop Quantum Gravity (LQG) \cite{livros1,livros3}, or more precisely in models inspired by LQG, in which the polymerisation of the classical Hamiltonian gives origin to (classically) empty, non-singular, spherically symmetric solutions with horizons \cite{modesto,corichi,AOS,mariam,alemaes2,guillermo2,espanhois,Rovelli3,bojowald}, sourced by a diffuse, effective energy-momentum tensor of quantum origin. The classical singularity is replaced by a transition surface from a black hole to a white hole. Among these models, we can find a particularly simple and reliable, with a single polymerisation parameter, symmetric black and white phases of same mass, asymptotic flat exterior metric, transition surface always inside the event horizon, and horizons with the same classical geometry \cite{espanhois}.

The radius of the transition surface can be fixed by imposing the constraint that its area equals the LQG minimum area \cite{modesto}. For the above model this was performed in \cite{Fernando,Fernando2}, where the model was also extended to charged black holes and to Planck scale horizons. The minimal area condition leads to a dependence of the polymerisation parameter on the black hole mass, with the former increasing as the latter decreases. It also implies a minimal mass for our solutions, with $m \geq \sqrt{2}/4$ in Planck units. It leads as well to the existence of extremal horizons, with masses $\sqrt{2}/4 \leq m \leq \sqrt{2}/2$ and charges $0 \leq Q \leq \sqrt{2}/2$, for which the surface gravity and temperature are zero. They constitute remnant black holes, asymptotic relics of Hawking evaporation \cite{Fernando2}. The role of Planck size remnants is particularly relevant in Cosmology, where they are natural candidates to compose dark matter \cite{Nature,modesto2,PLB,Nelson,referee1,theodor,Alejandro,Rovelli}.

Nevertheless, if there is no black hole solution with $m < \sqrt{2}/4$, what is there? Is the ``mass without mass" proposal realisable in this mass range? The purpose of the present paper is to find solutions in the context of the above polymer model, again under the assumption of a minimal area constraint. The latter will imply, once more, the presence of a minimal radius of the order of the Planck length. The resulting 
metric is static and asymptotically flat, but with no horizon. Diversely from the extremal horizons found in \cite{Fernando2}, which include neutral black holes, these non-singular naked solutions are always charged.

\section{Black and white holes}

The classical Reissner-Nordstr\"om Hamiltonian \cite{bascos3,rakesh,esteban} can be written as
\begin{equation}\label{eq:hcl}
    H_{\rm cl}=-\frac{1}{2 G \gamma}\left[\left(b+\frac{\gamma^2}{b}-\frac{\gamma^2Q^2}{b p_c} \right) p_b + 2c p_c    \right],
\end{equation}
where the conjugate variables obey the algebra $\{b,p_b\} = G\gamma$ and $\{c,p_c\} = 2G\gamma$ \cite{AOS}, $G$ is the gravitational constant and $\gamma$ is the Barbero-Immirzi parameter.
The internal, homogeneous metric assumes the form
\begin{equation} \label{homogeneous}
    ds^2 = - N^2 dT^2 + \frac{p_b^2}{p_c} dx^2 + p_c d\Omega^2,
\end{equation}
with a lapse $N = \gamma \sqrt{p_c}/b$. This metric can be quantised through a polymerisation procedure, defined in the present model by the transformations \cite{florencia}
\begin{equation}\label{eq:polymerisation}
b \rightarrow \dfrac{\sin(\delta_b b)}{\delta_b}, \quad \quad \quad p_b \rightarrow \dfrac{p_b}{\cos(\delta_b b)},
\end{equation}
where $\delta_b$ is the polymerisation parameter, equal to zero in the classical limit. In this way we obtain, after regularisation, the effective Hamiltonian
\begin{align}\label{eq:abbvhamiltonian}
H_{\rm eff}
&= -\dfrac{1}{2G\gamma\sqrt{1+\gamma^2\delta_b^2}}\left[  \left(\dfrac{\sin(\delta_b b)}{\delta_b} +\frac{\gamma^2 \delta_b}{\sin{(\delta_b b)}}- \dfrac{\gamma^2 \delta_b Q^2}{\sin( \delta_b b)p_c }\right)p_b + 2cp_c\cos(\delta_b b)\right].
\end{align}

Integrating the dynamical equation for $b$, it can be shown \cite{Fernando2} that $m$ defined by
\begin{equation}\label{eq:b}
    \frac{\sin^2(\delta_b b)}{\gamma^2 \delta_b^2}=\frac{2m}{\sqrt{p_c}}-1-\frac{Q^2}{p_c}
\end{equation}
is a constant of motion that will be identified with the black hole mass, whereas the equation for $p_c$ gives
\begin{equation}\label{eq:pc}
    p_c(T)=\frac{e^{-2 T}}{{4 b_0^4 (b_0+1)^2  m^2}}{\Bigl\{(b_0+1 )m^2 \bigl[ b_0-1 + ( b_0+1) e^T\bigr]^2  -  \
(b_0-1 ) b_0^2 Q^2\Bigr\}^2},
\end{equation}
with
\begin{equation} \label{b0}
b_0 \equiv \sqrt{1+\gamma^2 \delta_b^2}.
\end{equation}
We obtain the radius of the transition surface by doing $\dot{p}_c=0$. For $b_0 = 1$ (the classical case) we have $p_{c}^{\rm min} = 0$, while for $b_0 > 1$ the only real solution requires the condition
\begin{equation} \label{Qmax}
|Q|\leq m \frac{\sqrt{b_0^2-1}}{b_0},
\end{equation}
and it is given by
\begin{equation} \label{r0}
    r_0 = \sqrt{p_{c}^{\rm min}}=\frac{(b_0^2-1)}{b_0^2}m \left(\sqrt{1-\frac{b_0^2 Q^2}{(b_0^2-1)m^2}} + 1 \right).
\end{equation}

Solving the remaining Hamilton equations we can derive, with a suitable change of variables, the homogeneous inner metric
\begin{equation} \label{metric}
    ds^2 = - \left( \frac{2m}{r} - \frac{Q^2}{r^2} -1 \right)^{-1} \left( 1 - \frac{r_0}{m} g(r) \right)^{-1} dr^2 + \left( \frac{2m}{r} - \frac{Q^2}{r^2} -1 \right) d\tau^2 + r^2 d\Omega^2,
\end{equation}
where $r \equiv \sqrt{p_c}$, $r_0$ is its minimal value, and
\begin{equation} \label{gr}
    g(r) = \frac{\frac{2m}{r} - \frac{Q^2}{r^2}}{1 + \sqrt{1 - \frac{b_0^2 Q^2}{(b_0^2-1)m^2}}}.
\end{equation}
By analytic continuation, we obtain the external static metric
\begin{equation} \label{metric2}
    ds^2 = -\left( 1 - \frac{2m}{r} + \frac{Q^2}{r^2} \right) d\tau^2 + \left( 1 - \frac{2m}{r} + \frac{Q^2}{r^2} \right)^{-1} \left( 1 - \frac{r_0}{m} g(r) \right)^{-1} dr^2 + r^2 d\Omega^2,
\end{equation}
which is asymptotically flat and presents the same horizons as in the classical theory, with radii
\begin{equation} \label{horizons}
    r_h^{\pm} = m \left( 1 \pm \sqrt{1 - \frac{Q^2}{m^2}} \right).
\end{equation}

\section{Stable horizons}

The quantity $m$ is an integration constant in (\ref{eq:b}), which can be fixed by imposing the minimal area condition
\begin{equation} \label{minimal_area}
    4\pi p_{c}^{\rm min} = 4 \pi \sqrt{3} \gamma,
\end{equation}
which leads to\footnote{We will take $\gamma = \sqrt{3}/6$, the value obtained in \cite{CQG} from the entropy analysis of quantum horizons, but our conclusions do not depend on this choice.}
\begin{equation} \label{delta_b}
    \delta_b^2 = \frac{12}{2\sqrt{2}m - 2Q^2 -1}.
\end{equation}
As $m$ and $Q$ are constants of motion, $\delta_b$ is also a constant of motion along a given dynamical trajectory, as assumed when the Hamilton equations were derived from (\ref{eq:abbvhamiltonian}). This can be achieved by promoting $\delta_b$ as the conjugate of a cyclic dynamical variable \cite{AOS}. The Hamiltonian presents a wider set of solutions, but the constraint (\ref{delta_b}) selects only those that match the LQG minimal area (\ref{minimal_area}) at the transition surface.

Combining (\ref{delta_b}) and (\ref{Qmax}) leads to an upper limit for the black hole charge,
\begin{equation} \label{Qmax2}
  Q^2 \leq \frac{\sqrt{2}}{2} m.
\end{equation}
From (\ref{delta_b}) we also have
\begin{equation} \label{limite}
    2\sqrt{2}m > 2Q^2 + 1,
\end{equation}
which, from (\ref{horizons}), implies that
\begin{equation}
    r_h^- < r_0 < r_h^+,
\end{equation}
that is, the Cauchy horizon lies beyond the transition surface and is not reached.
When the inequality (\ref{limite}) is saturated, $\delta_b$ diverges, $b_0 \gg 1$ and $r_h^+ \rightarrow r_0$, whereas
\begin{equation}
    r_h^- = 2m - \frac{\sqrt{2}}{2}.
\end{equation}
For $Q=0$ we have $r_h^- = 0$ and $r_h^+/2 = m = \sqrt{2}/4$, while for the maximal charge (\ref{Qmax2}) one has $r_h^{\pm} = Q = m = \sqrt{2}/2$.

The Komar energy \cite{wald} is given by
\begin{equation} \label{Komar}
    E_K(r) = \left( m - \frac{Q^2}{r} \right) \sqrt{1 - \frac{r_0}{m} g(r)}.
\end{equation}
For an asymptotic observer we have $E_K = m$, while at the transition surface $E_K(r_0)=0$. That is, when (\ref{limite}) is saturated, the Komar energy at the event horizon vanishes, and so the horizon temperature.

\section{Naked solutions}

When condition (\ref{limite}) is not satisfied, we have $\delta_b^2 < 0$ and, doing $\delta_b = i \tilde{\delta}_b$, the polymerisation (\ref{eq:polymerisation}) assumes the form
\begin{equation}\label{eq:polymerisation2}
b \rightarrow \dfrac{\sinh(\tilde{\delta}_b b)}{\tilde{\delta}_b}, \quad \quad \quad p_b \rightarrow \dfrac{p_b}{\cosh(\tilde{\delta}_b b)},
\end{equation}
where, now,
\begin{equation} \label{delta_b2}
    \tilde{\delta}_b^2 = \frac{12}{1 + 2Q^2 - 2\sqrt{2}m},
\end{equation}
and
\begin{equation} \label{b02}
b_0 = \sqrt{1-\gamma^2 \tilde{\delta}_b^2}.
\end{equation}
In the presence of horizons, the polymerisation (\ref{eq:polymerisation2}), with periodic functions replaced by hyperbolic ones, is the same used to obtain external metrics like (\ref{metric2}), which can be derived from the inner metric (\ref{homogeneous}) through the transformations $b \rightarrow ib$ and $p_b \rightarrow i p_b$ \cite{AOS,Fernando}. 

Now we have the opposite, and the external metric (\ref{metric2}) is found through the polymerisation (\ref{eq:polymerisation}) (with $\delta_b \rightarrow \tilde{\delta}_b$).  
There is again a minimum radius $r_0 = \sqrt{2}/2$, root of $\dot{p}_c = 0$. The minimum occurs for
\begin{equation}
    e^{2T} = \left( \frac{b_0-1}{b_0+1} \right)^2 - \frac{(b_0-1)b_0^2Q^2}{(b_0+1)^3m^2}.
\end{equation}
As $T$ is a real coordinate, for $b_0>1$ the minimum (\ref{r0}) exists only if condition (\ref{Qmax}) is fulfilled, as already seen. On the other hand, for $b_0<1$
the only real root is expressed as
\begin{equation} \label{r02}
   r_0 = \sqrt{p_{c}^{\rm min}}=\frac{(1-b_0^2)}{b_0^2}m \left(\sqrt{1+\frac{b_0^2 Q^2}{(1- b_0^2)m^2}} -1 \right),
\end{equation}
while $g(r)$ in (\ref{metric2}) assumes the form
\begin{equation} \label{gr2}
    g(r) = \frac{\frac{2m}{r} - \frac{Q^2}{r^2}}{1 - \sqrt{1 + \frac{b_0^2 Q^2}{(1-b_0^2)m^2}}}.
\end{equation}
The condition $b_0^2 > 0$ requires the constraint
\begin{equation} \label{Qmax5}
    Q^2 >
 \sqrt{2} m.
\end{equation}

\begin{figure}[!t]
     \centering
     \begin{subfigure}[b]{0.45\textwidth}
         \centering
         \includegraphics[width=\textwidth]{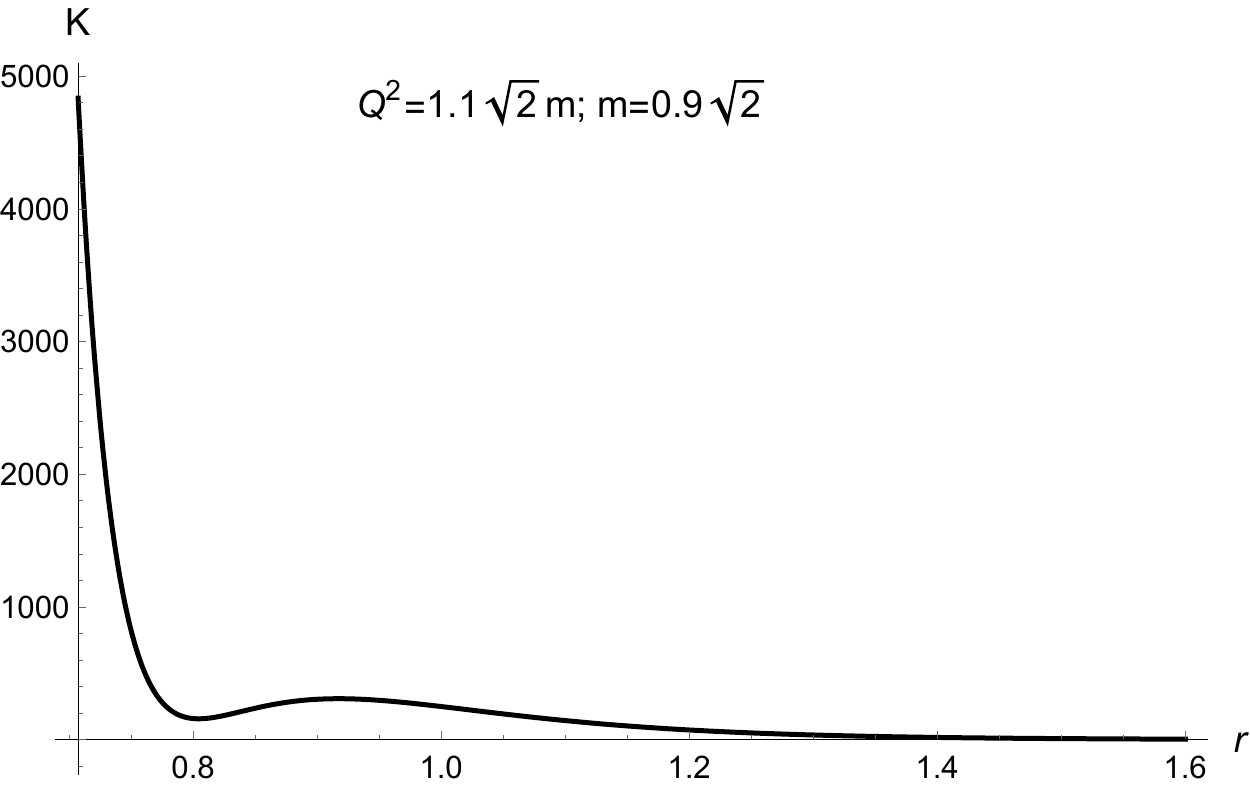}
         \label{fig:Qzero}
     \end{subfigure}
     \hfill
     \begin{subfigure}[b]{0.45\textwidth}
         \centering
         \includegraphics[width=\textwidth]{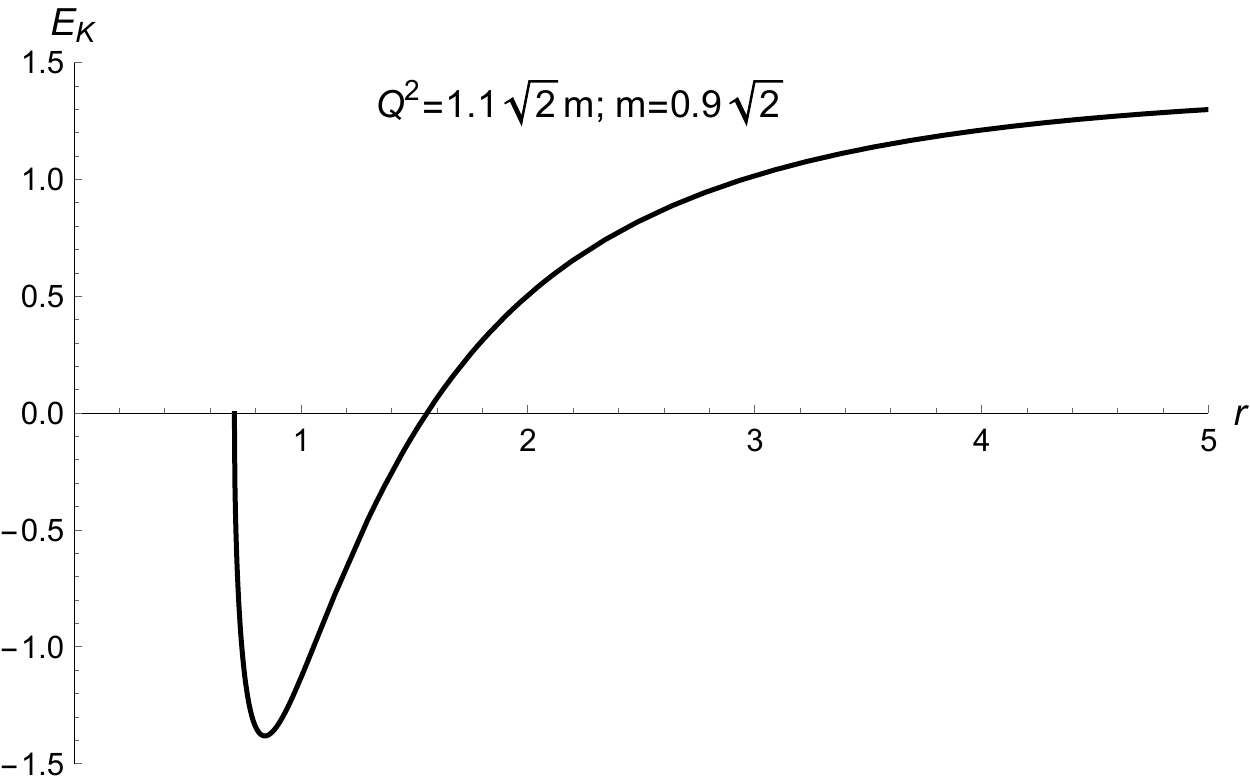}
         \label{fig:Qmax}
     \end{subfigure}
     \hfill
        \caption{Kretschmann scalar (left) and Komar energy (right) for the naked metric, as functions of the radial coordinate.}
        \label{fig:Kretschmann}
\end{figure}

From (\ref{Qmax5}) and (\ref{horizons}) we can see that, for $m < \sqrt{2}$, metric (\ref{metric2}) presents no horizon, representing a non-singular naked solution. The spacetime is complete, as for $-\infty < T < \infty$ the metric covers (twice) the range $r \geq r_0$. 
Note that, under condition (\ref{Qmax5}), 
\begin{equation} \label{g}
    1 - \frac{r_0}{m} g(r) > 0
\end{equation}
for $r > r_0$, which means that metric (\ref{metric2}) is regular everywhere, as it is also evident from the left panel of Fig.~1, where we show the Kretschmann scalar for $r \geq r_0$.
The result (\ref{g}) also guarantees that the Komar energy (\ref{Komar}) is real everywhere. It is equal to $m$ for an asymptotic observer, but, as shown in the right panel of Fig.~1, becomes negative for $r < Q^2/m$. Notice as well that $g_{rr}$ diverges at $r = r_0$, where the left-hand side of (\ref{g}) is zero. It is not a physical singularity as the curvature is finite. It only means that no space is defined beyond this radius, where $p_c(T)$ presents a minimum and null geodesics are bounced. Indeed, by doing $ds^2 = 0$ in (\ref{metric2}) we obtain
\begin{equation}
    \frac{dr}{d\tau} = \left( 1 - \frac{2m}{r} + \frac{Q^2}{r^2} \right) \sqrt{1 - \frac{r_0}{m} g(r)},
\end{equation}
which vanishes at $r=r_0$.

It is possible to show that the minimal radius is inaccessible for a classical, neutral test particle initially at rest at infinity, because its geodesic suffers a bounce at $r > r_0$. Its proper velocity $v = dl/dt$, i.e. measured by a local inertial observer, is given by
\begin{equation}
    v^2 = \left(\frac{2m}{r} - \frac{Q^2}{r^2} \right),
\end{equation}
where $dl^2 = g_{rr} dr^2$ and $dt^2 = -g_{00} d\tau^2$. We see that it becomes zero for $r = Q^2/(2m) > r_0$. Incidentally, the turnover occurs inside the region of negative Komar energy. In fact, the proper acceleration is expressed as
\begin{equation}
    \frac{dv}{dt} = -\left(1 - \frac{r_0}{m} g(r) \right)^{1/2} \left( \frac{m}{r^2}-\frac{Q^2}{r^3} \right) \left(1 - \frac{2m}{r} + \frac{Q^2}{r^2} \right)^{1/2},
\end{equation}
and it is inverted for $r = Q^2/m$, when the Komar energy becomes negative.

Evidently, particles with high enough kinetic energy could approach the minimum radius as close as desired. However, no reliable observation of such an object could involve an exchange of momenta larger than $m$ and, therefore, distance scales below $1/m$ could not be resolved. As our naked solutions correspond to $m < \sqrt{2}$, this means that the minimal radius $r_0 = \sqrt{2}/2$ is not accessible. In the same way, for charges below the unity, the region with negative Komar energy cannot be observed.




The gravitational repulsion near the minimal radius can also be understood in terms of the effective energy density and pressures associated to the spacetime fluctuations. The Einstein equations for the static metric (\ref{metric2}) are reduced to \cite{Landau}
\begin{equation}
    8\pi \rho = - e^{-\lambda} \left( \mu'' + \frac{3}{4} \mu'^2 - \frac{\mu' \lambda}{2} \right) + e^{-\mu},
\end{equation}
\begin{equation}
    8\pi p_r = \frac{e^{-\lambda}}{2} \left( \frac{\mu'^2}{2} + \mu' \nu' \right) - e^{-\mu},
\end{equation}
\begin{equation}
    8\pi p_t = \frac{e^{-\lambda}}{4} \left( 2\nu'' + \nu'^2 + 2 \mu'' + \mu'^2 - \mu' \lambda' - \nu'\lambda' + \mu'\nu'\right),
\end{equation}
where $\rho$, $p_r$ and $p_t$ are, respectively, the effective energy density, the radial and the tangential pressures, the prime means derivative with respect to $r$, and we defined $e^{\mu} = r^2$,
\begin{equation}
    e^{\nu} = 1 - \frac{2m}{r} + \frac{Q^2}{r^2},
\end{equation}
\begin{equation}
    e^{\lambda} = e^{-\nu} \left( 1 - \frac{r_0}{m} g(r) \right)^{-1}.
\end{equation}

\begin{figure}[t]
     \centering
     \begin{subfigure}[b]{0.45\textwidth}
         \centering
         \includegraphics[width=\textwidth]{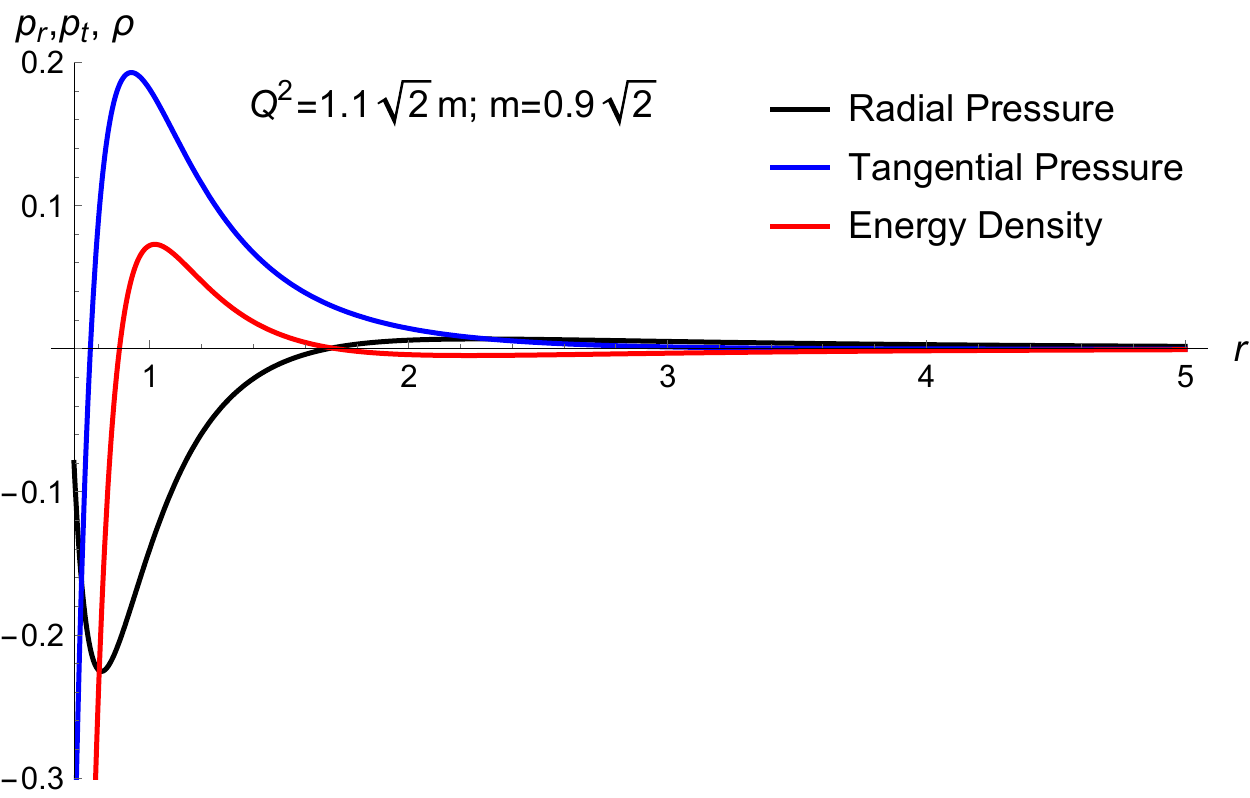}
         \label{fig:Qzero2}
     \end{subfigure}
     \hfill
    \begin{subfigure}[b]{0.45\textwidth}
         \centering
         \includegraphics[width=\textwidth]{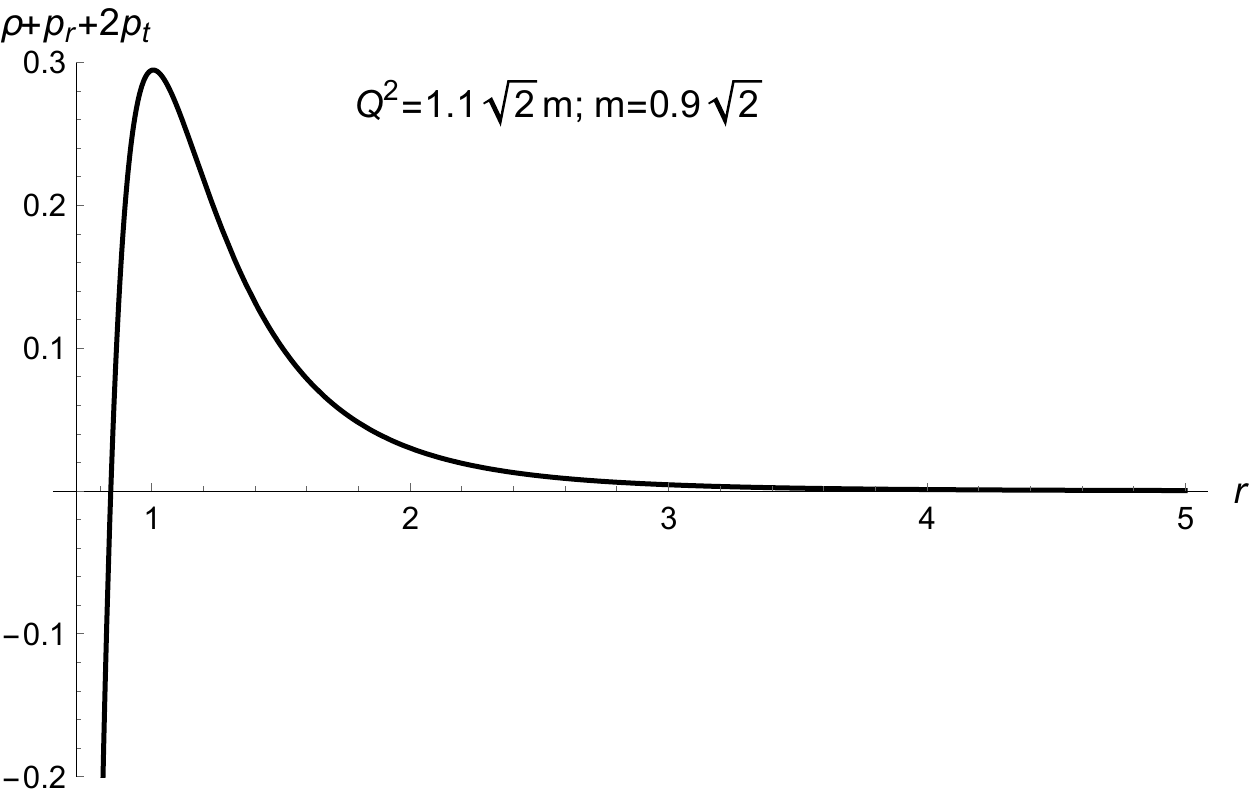}
         \label{fig:Qzero3}
     \end{subfigure}
     \hfill
        \caption{Effective pressures and energy density for the naked metric as functions of the radial distance (left). In the right panel the combination $\rho + p_r + 2p_t$ is also shown.}
        \label{fig:pressure_energy}
\end{figure}

In the left panel of Fig.~2 we give an example of the radial variation of the effective pressures and energy density, while in the right panel we show the combination $\rho + p_r + 2p_t$, which turns out to be negative for short distances, violating the strong energy condition. 
The energy conditions should, however, be interpreted with caution as the Einstein equations are not valid in this context, where $\rho$, $p_r$ and $p_t$ are effective quantities.


\section{Concluding remarks}

We had previously shown that stable horizons only exist for masses between $m = \sqrt{2}/4$ (the uncharged solution) and $m = \sqrt{2}/2$, when the charge has its maximum value $Q = \sqrt{2}/2$. On the other hand, we have seen in the present work that, for this maximal charge, the naked solution has a maximum mass $m = \sqrt{2}/4$. This mass represents, therefore, a threshold above which non-singular (neutral or charged) black-holes can be formed, while for masses below this limit we have only non-singular naked solutions. 

The case with horizons can be seen as realisation of the ``mass without mass" proposal, in the sense that we have empty solutions sourced by spacetime quantum fluctuations, although the addition of matter fields is always possible. The existence of non-singular naked solutions with masses below the Planck scale might suggest a description of elementary particles as configurations of spacetime as well. However, this would be difficult to implement due to the presence of weak and strong interactions in the standard model, which would require considering non-abelian solutions, with a symmetry structure far from simple. Not to mention the difficulties involved in describing spin-carrying particles and in realising the ``charge without charge" proposal as in the original conception by Misner and Wheeler.

Anyway, it seems promising that quantum theories of spacetime, even in effective models like this, may not only avoid the singular fate of black holes but also allow the existence of non-singular ``naked singularities", overcoming in this way the cosmic censor conjecture.

\section*{Acknowledgements}

We are thankful to A. Saa, D. Cook, F. G. Menezes and A. Olimpieri for helpful discussions. SC is partially supported by CNPq (Brazil) with grant 308518/2023-3.

\end{document}